\documentstyle[psfig]{l-aa}

\begin{document}

   \thesaurus{04         
              (
               08.01.1;  
               08.06.3;  
               10.01.1;  
               10.05.1   
              )} 
 
   \title{
          The rise and fall of the NaMgAl stars
   \thanks{
           Table 4 is only available in electronic form at the CDS via
           anonymous ftp to cdsarc.u-strasbg.fr (130.79.128.5) or via
           http://cdsweb.u-strasbg.fr/Abstract.html.
          }}

   \author{
           Jocelyn Tomkin\inst{1}
   \and    Bengt Edvardsson\inst{2}
   \and    David L. Lambert\inst{1}
   \and    Bengt Gustafsson\inst{2}
          }

   \offprints{J. Tomkin}

   \institute{
              McDonald Observatory and Department of Astronomy, University of
              Texas, Austin, Texas 78712, U.S.A.
   \and       Uppsala Astronomical Observatory, Box 515, S-751 20 Uppsala,
              Sweden
             }

   \date{Received 20 May 1997 / Accepted 19 June 1997}

   \maketitle

   \markboth{J. Tomkin et al.: NaMgAl stars}
            {J. Tomkin et al.: NaMgAl stars}

   \begin{abstract}

We have made new abundance determinations for a sample of NaMgAl stars.
These stars, which are a subgroup of the nearby metal-rich field F and G
disk dwarfs, were first identified by Edvardsson et al. (1993) on the
basis of their apparent enrichment in Na, Mg and Al relative to other
elements.  The discovery of a planetary companion to the nearby solar type star
51~Peg (Mayor \& Queloz 1995) combined with Edvardsson et al.'s earlier
identification of 51~Peg as a NaMgAl star highlighted
the group's potential importance.  Our new analysis, which uses new spectra
of higher resolution and better wavelength coverage than the analysis of
Edvardsson et al., shows that the Na, Mg and Al abundances of the
NaMgAl stars are indistinguishable from those of non-NaMgAl stars
with otherwise similar properties.  The group thus appears to be spurious.

Our study, which includes 51~Peg, also provides the most
complete set of abundances for this star available to date.  The new 
Fe abundance, ${\rm [Fe/H]} = +0.20 \pm 0.07$, of 51~Peg confirms
earlier measurements of its metal richness.  Abundances for 19 other
elements, including C, N and O, reveal a fairly uniform enrichment
similar to that of Fe and show no evidence of abnormality
compared to other metal rich stars of similar spectral type.

   \end{abstract}


   \section{Introduction}

In a survey of abundances in nearby field F and G dwarfs
Edvardsson et al. (1993, EAGLNT below) suggested the existence of a new group of
stars that are enriched in Na, Mg and Al relative to other elements.
EAGLNT identified eight of these ``NaMgAl'' stars in their survey of 189 stars.
The average enhancements relative to other stars of similar metallicity are
$\Delta {\rm [X/Fe]} = 0.12, 0.14, 0.10$ for Na, Mg and Al, respectively.
\footnote{In this paper we use two customary spectroscopic notations:
[X/Y] $\equiv \log_{\rm 10}$(X/Y)$_{\rm star} - \log_{\rm 10}$(X/Y)$_{\odot}$
to define the logarithmic abundances relative to the Sun,
and $\log \varepsilon$(X) $\equiv \log_{\rm 10}$(X/H)$ + 12$}
These stars are also characterized by metal richness -- all have
[Fe/H]~$ > 0.05$.  EAGLNT suggested that perhaps ``the stars have been
affected by mass transfer from a now `dead' companion''.  The
recent discovery of planetary companions to nearby solar
type stars has given this speculation unexpected immediacy.

The first of these discoveries was 51~Peg (Mayor \& Queloz 1995).
Mayor and Queloz's radial velocity observations 
revealed the presence of a companion with a minimum mass
${\cal M} \sin i = 0.47 \pm 0.02 {\cal M}_{\rm J}$
(${\cal M}_{\rm J}=$ the mass of Jupiter),
a result confirmed by Marcy et al. (1997).
A remarkable feature of the companion is its
short orbital period -- only 4.2 days.  This places the 
companion only 0.05\,au from its parent star, which is 
puzzlingly close for a planet of Jupiter-type mass.  51~Peg
was included in EAGLNT's survey and labelled as a NaMgAl star.  The
suggestion as to these stars' origin thus raises the
possibility that 51~Peg's companion may be an
exotic form of stellar remnant, rather than a planet.

Further discoveries of planetary companions have followed 51~Peg.
Three of these additional new systems -- $\upsilon$~And, $\tau$ Boo 
and $\rho^1$~Cnc (Butler et al. 1997) -- mimic 51~Peg
in that they are solar type stars and their companions are very close
to their parent stars.  Two of these systems ($\tau$~Boo and 55~Cnc)
were not included in the survey of EAGLNT, while the
third ($\upsilon$~And) was, but was not identified as a NaMgAl
star.  This extremely limited sample thus suggests that the
possible correlation between the NaMgAl stars and close planetary
companions is imperfect, nonetheless
it is clear that the NaMgAl stars deserve closer examination.  
\begin{table*}  
\caption[]{
Programme star data. Column 3) indicates whether the star was
identified as a NaMgAl star in EAGLNT. Columns 4)--7) show the model atmosphere
parameters, 8)--10) show velocities relative to a local standard of rest where
the $U$ velocity is taken positive towards the galactic centre.
Columns 11)--15) give kinematic data and ages adopted from EAGLNT
}
\begin{flushleft}  
\begin{tabular}{rrcrrrrrrrrrrrrr} 
\noalign{\smallskip}
\hline
\noalign{\smallskip}
HR & Star & NaMgAl & $T_{\rm eff}$ & $\log g$ & [M/H] & $\xi_{\rm t}$ & $U$ & $V$ & $W$ &\ &
 $R_{\rm p}$ & $R_{\rm m}$ & $Z_{\rm max}$ & $e$~~~ & $\log {\rm Age\over 10^9 yr}$ \\
\cline{7-10}\cline{12-14}
   &      &     & [K]           & [cgs]    &       & \multicolumn{4}{c}{[km\,s$^{-1}$]} &\ &
 \multicolumn{3}{c}{[kpc]} \\
\hline
 448 & --        & yes & 5825 & 4.10 & 0.10 & 1.60 &     0 & $-$25 &    19 &\ & 6.6 & 7.3 & 0.20 & 0.10 & 0.53 \\ 
1536 & --        & yes & 5930 & 4.20 & 0.20 & 1.55 &    41 & $-$80 & $-$12 &\ & 3.8 & 6.0 & 0.10 & 0.36 & 0.64 \\ 
3176 & $\mu$ Cnc & no  & 5835 & 4.10 & 0.10 & 1.75 & $-$42 &    10 & $-$10 &\ & 7.4 & 8.6 & 0.10 & 0.14 & 0.94 \\ 
3951 & 20 LMi    & yes & 5800 & 4.40 & 0.20 & 1.25 &    61 & $-$62 &    11 &\ & 4.5 & 5.5 & 0.09 & 0.31 & 0.81 \\ 
4027 & 24 LMi    & no  & 5875 & 4.15 & 0.05 & 1.50 &    11 & $-$16 &    24 &\ & 7.0 & 7.6 & 0.26 & 0.08 & 0.79 \\ 
4688 & 9 Com     & yes & 6340 & 4.20 & 0.20 & 2.10 &    12 & $-$35 &  $-$7 &\ & 6.0 & 7.0 & 0.06 & 0.14 & 0.32 \\ 
8041 & 11 Aqr    & no  & 5920 & 4.40 & 0.25 & 1.40 &  $-$3 & $-$23 &  $-$1 &\ & 6.7 & 7.3 & 0.01 & 0.09 & 0.89 \\ 
8472 & --        & yes & 6325 & 3.95 & 0.10 & 2.35 &    29 & $-$22 &     0 &\ & 6.5 & 7.4 & 0.00 & 0.13 & 0.40 \\ 
8729 & 51 Peg    & yes & 5775 & 4.35 & 0.20 & 1.25 &    12 & $-$24 &    18 &\ & 6.6 & 7.4 & 0.19 & 0.10 & 0.93 \\ 
  -- & The Sun   & no  & 5780 & 4.44 & 0.00 & 1.15 & $-$10 &     6 &     6 &\ & 7.9 & 8.4 & 0.06 & 0.06 & 0.66 \\
\hline
\end{tabular}
\end{flushleft}
\end{table*}

\begin{figure*}
\centerline{\psfig{figure=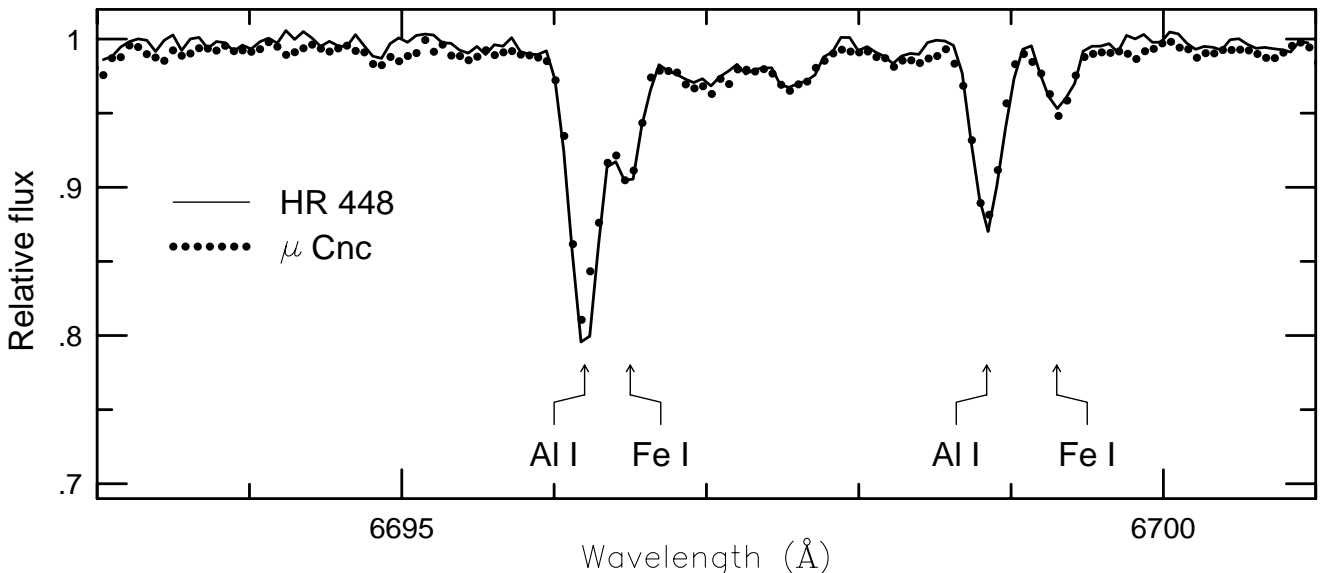,width=18cm}}
\caption[]{
Two Al\,{\sc i} and two Fe\,{\sc i} lines near 6700\,\AA\ in
HR\,448, a NaMgAl star, and $\mu$ Cnc, a non-NaMgAl star.  The
spectrum of $\mu$ Cnc has been wavelength shifted to match that
of HR\,448.  The two stars have similar effective temperatures
and gravities (see Table~1) so the striking similarity of their spectra
indicates that any enhancement of Al relative to Fe in HR\,448
is very small
}
\end{figure*}

Here we report an abundance analysis of the
NaMgAl stars based on new observations made with the
McDonald Observatory's 2.7\,m telescope and 2dcoud\'e
spectrometer.  The generous wavelength coverage of this
instrument provides several more lines of the three
elements than EAGLNT were able to use; their
observations only covered a few limited wavelength regions.
And, for the northern hemisphere stars in the EAGLNT
survey, the high resolution ($R = $\,60\,000) of this
instrument also provides more accurate equivalent widths.
The high resolution is particularly significant for analysis of metal-rich stars
like 51~Peg.
(The northern hemisphere stars of the survey of EAGLNT, were observed with a
resolution of 30\,000, while 
those of the southern hemisphere were observed with
resolutions of 60\,000 or 80\,000).
The wide wavelength coverage also allows access to some additional
elements not included previously.

The main objective of the new study of the NaMgAl stars,
therefore, is to inspect the enhancements of Na, Mg and Al,
which are small, more closely and to see if any other elements
are involved.  A secondary objective is to expand the set of
elements with measured abundances for 51~Peg.


   \section{Observations}

The observations were made at McDonald Observatory with the 2.7\,m
telescope and 2dcoud\'e echelle spectrometer (Tull et al. 1995).
The detector was a Tektronix CCD with 24\,$\mu$m$^2$ pixels
arranged in a 2048$\times$2048 pixel format.

We observed six of the eight NaMgAl stars identified by
EAGLNT and three other stars in the EAGLNT
survey not identified as NaMgAl stars, but with
similar effective temperatures, gravities and
metallicities.  These nine stars, which include 51~Peg, are given
in Table~1.  Finally, we observed an asteriod (Iris) in order to
provide a solar spectrum recorded under the same circumstances as the
stellar spectra.

We observed the wavelength interval 4000 -- 9000\,\AA\ approximately.
The coverage is complete from the start of this interval to
5600\,\AA\ and substantial, but incomplete, from 5600\,\AA\ to
the end of the interval because of gaps between the end of one
echelle order and the beginning of the next.  The slit width
was set to project onto two pixels, which gave a resolution of
60\,000.  From $\sim5500$ to $\sim9000$\,\AA\ the extracted
one-dimensional stellar spectra have a typical signal-to-noise
ratio of $\sim400$, while at shorter wavelengths than 
$\sim5500$\,\AA\ the signal-to-noise ratio decreases with
decreasing wavelength because of the decline of
the stellar (and flat field) fluxes towards shorter wavelengths.

Figure~1 shows the two Al\,{\sc i} lines near 6700\,\AA\ in a NaMgAl star
and a non-NaMgAl star.  The two stars have similar effective temperatures
and gravities so differences in the strengths of their lines primarily
reflect abundance differences.  The strengths of the Al\,{\sc i} lines
relative to the Fe\,{\sc i} lines in the two stars are very similar
suggesting that any enhancement of Al relative to Fe in the NaMgAl
star is small.

The data were processed and wavelength calibrated in a
conventional manner with the IRAF package of programs on a
SPARC 5 workstation in the Astronomy Department at the
University of Texas.  Lines suitable for measurement were chosen
for clean profiles, as judged by inspection of the solar
spectrum at high resolution and signal-to-noise ratio
(Kurucz et al. 1984), that provide reliable equivalent
widths in all, or most, of the programme stars.  Moore et al. (1966)
was our primary source of line identification.  The
equivalent width of each line was measured with the IRAF
measurement option most suited to the situation of the line;
usually this was the fitting of a single, or multiple,
Gaussian profile to the line profile.

Table~2 gives the list of lines measured.  The list includes
most of the lines used by
EAGLNT and many additional ones.
In particular our new list contains more Na, Mg and Al lines
than theirs; specifically three Na\,{\sc i},
four Mg\,{\sc i} and seven Al\,{\sc i} lines in the new
line list compared with two Na\,{\sc i}, two Mg\,{\sc i}
and two Al\,{\sc i} lines in theirs.  The new 
line list also has more lines of most other elements and it
includes some new elements, notably C and N which are
represented by six C\,{\sc i} lines and two N\,{\sc i} lines.

Figure~2 compares our new equivalent widths with those of EAGLNT.
The overall agreement is good.
In a few cases, however, we note a tendency for the new equivalent widths to be
slightly larger than those of EAGLNT; 51~Peg is the most marked case.
This can be attributed to the fact that for most of the stars in the
programme the resolution (60\,000) of the new observations
is higher than that (30\,000) of EAGLNT's observations
thus leading to a better defined and slightly higher continuum.
\footnote{Of the nine stars in our programme seven ($\mu$~Cnc,
20~LMi, 24~LMi, 9~Com, 11~Aqr, HR~8472, 51~Peg) had been
observed from McDonald Observatory at a resolution of 30\,000
and two (HR~448, HR~1536) had been observed from ESO at a
resolution of 80\,000 in EAGLNT's study.}
The behaviour of 51~Peg, which as one of the coolest and most
metal rich of the stars has one of the most crowded spectra,
supports this explanation.


\begin{figure}
\centerline{\psfig{figure=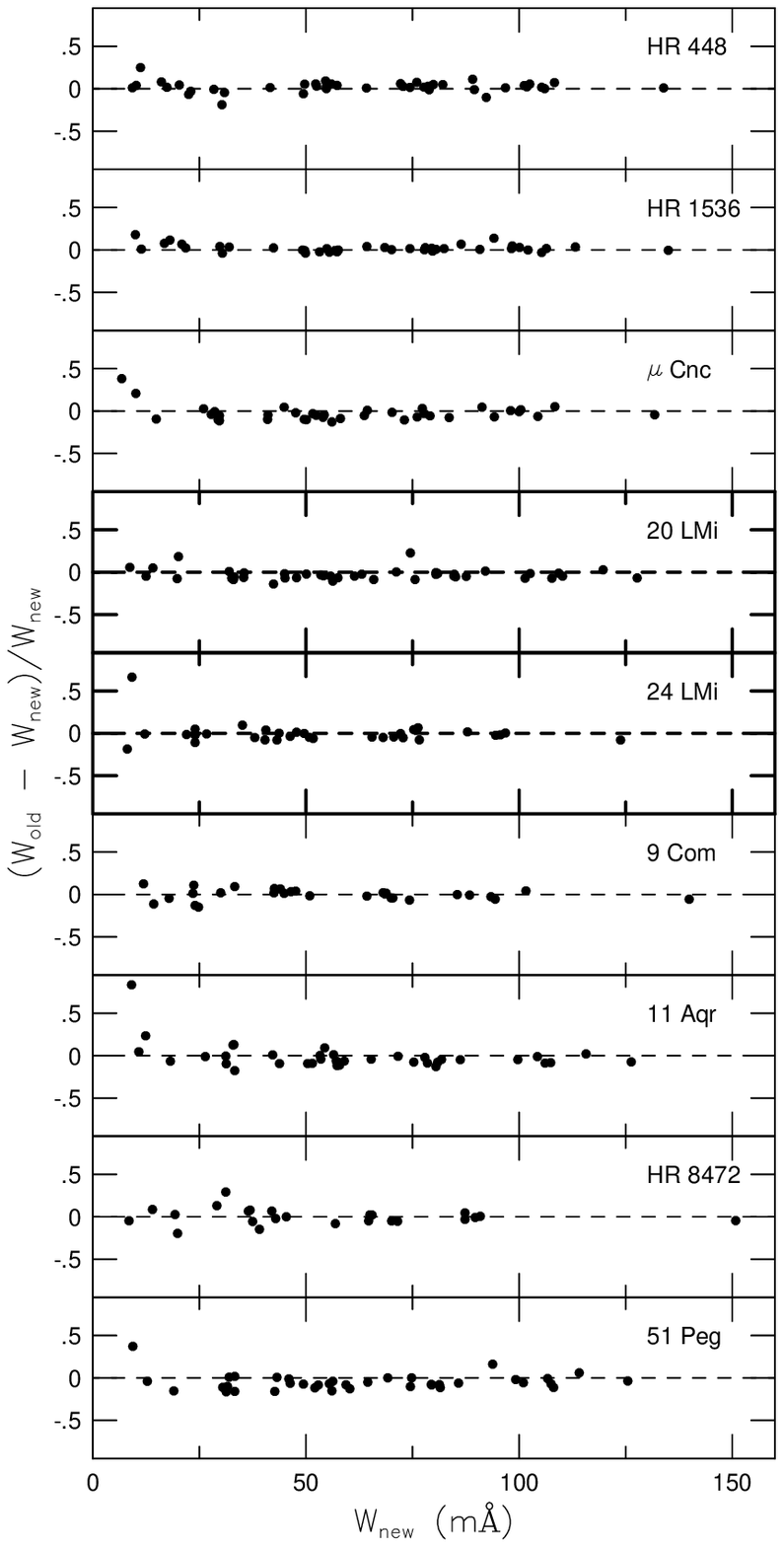,width=9cm}}
\caption[]{
Comparison of stellar equivalent widths with those of EAGLNT
}
\end{figure}
   \section{Analysis}

The LTE abundance analysis was made strictly relative to the Sun and adopts the
same procedures, software and model atmospheres as were used by EAGLNT.

   \subsection{Line data}

Basic line data for all the measured transitions were obtained from the VALD
atomic line data base (Piskunov et al. 1995) by the ``show line'' command.
Astrophysical oscillator strengths were determined for each line by requiring
that the equivalent width derived from the solar model be equal to that measured
from the solar flux spectrum as reflected by Iris.
The parameters of the solar model are given in Table~1
and the solar chemical abundances were adopted from Grevesse et al. (1996).
New quantum
mechanical calculations of pressure line-broadening properties by
Anstee \& O'Mara (1995) and Barklem \& O'Mara (1997) were adopted for several
lines of s-p, p-s, p-d and d-p transitions in neutral species.
There is a small systematic difference between our solar equivalent widths and
those measured by EAGLNT.
The weakest lines are in the mean about 2\% stronger in EAGLNT, and the
strongest lines about 2\% weaker.
This difference/uncertainty (together with the line-broadening mentioned above)
is reflected in the $gf$ values.
However, since the programme stars are observed and measured with identical
procedures, we neglect the very small differential errors which may be caused by
these uncertainties.
\setcounter{table}{1}
\begin{table}
\caption[]{
Atomic line data and solar equivalent widths.
The columns give: 1) the line rest wavelength, 2) the excitation
energy of the lower level of the transition, 3) the logarithm of the product
of the statistical weight and the oscillator strength, 4) an enhancement
factor to the classical van~der~Waals damping constant; An ``A'' or a ``B''
indicates that the pressure damping was instead calculated according to
Anstee \& O'Mara (1995) or Barklem \& O'Mara (1997) respectively, 5) the
radiation damping constant, and 6) the equivalent width measured from the
solar spectrum as reflected by the asteroid Iris
}
\begin{flushleft}
\begin{tabular}{rrrcrr}
\hline\noalign{\smallskip}
 $\lambda$~~~~ & $\chi_{\rm l}$~ & $\log gf$ & $\delta\Gamma_6$ & $\Gamma_{\rm rad}$~~~ & W$_{\lambda\odot}$ \\
 $[$\AA]~~~    & [eV]            &           &                  &       s$^{-1}$~~~     & [m\AA ]            \\
\noalign{\smallskip}
\hline\noalign{\smallskip}
\multicolumn{4}{l}{{\bf C I}~~~ $\log\epsilon_\odot=8.55$} \\
   5380.32 &  7.69 & $-$1.68 & 2.5 & 1.7\,10$^6$ &  22.3 \\
   6587.62 &  8.54 & $-$1.14 & 2.5 & 3.9\,10$^6$ &  14.6 \\
   7111.45 &  8.64 & $-$1.18 & 2.5 & 1.8\,10$^6$ &  11.2 \\
   7113.17 &  8.65 & $-$0.78 &  B  & 2.5\,10$^6$ &  22.7 \\
   7115.17 &  8.64 & $-$0.68 &  B  & 4.0\,10$^6$ &  26.8 \\
   7116.96 &  8.16 & $-$1.30 & 2.5 & 1.4\,10$^6$ &  18.5 \\
\multicolumn{4}{l}{{\bf N I}~~~ $\log\epsilon_\odot=7.97$} \\
   7468.27 & 10.34 & $-$0.06 &  A  & 3.1\,10$^7$ &   4.1 \\
   8629.16 & 10.67 &    0.08 &  A  & 3.0\,10$^7$ &   3.4 \\
\multicolumn{4}{l}{{\bf O I}~~~ $\log\epsilon_\odot=8.87$} \\
   6158.17 & 10.74 & $-$0.37 &  B  & 5.4\,10$^7$ &   4.2 \\
   7771.95 &  9.14 &    0.36 &  A  & 4.8\,10$^7$ &  70.6 \\
   7774.18 &  9.14 &    0.21 &  A  & 4.7\,10$^7$ &  60.9 \\
   7775.40 &  9.14 & $-$0.02 &  A  & 4.5\,10$^7$ &  47.9 \\
\multicolumn{4}{l}{{\bf Na I}~~~ $\log\epsilon_\odot=6.33$} \\
   4497.68 &  2.10 & $-$1.59 & 2.0 & 6.4\,10$^7$ &  39.3 \\
   6154.23 &  2.10 & $-$1.59 & 2.0 & 6.4\,10$^7$ &  39.6 \\
   6160.75 &  2.10 & $-$1.32 & 2.0 & 6.7\,10$^7$ &  57.7 \\
\multicolumn{4}{l}{{\bf Mg I}~~~ $\log\epsilon_\odot=7.58$} \\
   6318.71 &  5.11 & $-$2.08 & 2.5 & 2.8\,10$^5$ &  39.8 \\
   6319.24 &  5.11 & $-$2.28 & 2.5 & 3.0\,10$^5$ &  27.8 \\
   6965.41 &  5.75 & $-$1.75 & 2.5 & 3.6\,10$^5$ &  24.1 \\
   7759.37 &  5.93 & $-$1.68 & 2.5 & 3.3\,10$^5$ &  21.0 \\
\multicolumn{4}{l}{{\bf Al I}~~~ $\log\epsilon_\odot=6.47$} \\
   6696.03 &  3.14 & $-$1.59 & 2.5 & 3.0\,10$^8$ &  38.1 \\
   6698.67 &  3.14 & $-$1.91 & 2.5 & 3.0\,10$^8$ &  21.9 \\
   7835.32 &  4.02 & $-$0.72 & 2.5 & 7.9\,10$^7$ &  48.1 \\
   7836.13 &  4.02 & $-$0.57 & 2.5 & 7.9\,10$^7$ &  61.0 \\
   8772.88 &  4.02 & $-$0.46 & 2.5 & 8.3\,10$^7$ &  75.2 \\
   8773.91 &  4.02 & $-$0.29 & 2.5 & 8.3\,10$^7$ &  94.4 \\
   8828.87 &  4.09 & $-$1.89 & 2.5 & 4.2\,10$^6$ &   4.0 \\
\multicolumn{4}{l}{{\bf Si I}~~~ $\log\epsilon_\odot=7.55$} \\
   6125.03 &  5.61 & $-$1.53 & 1.3 & 1.1\,10$^6$ &  33.4 \\
   6142.49 &  5.62 & $-$1.47 & 1.3 & 8.8\,10$^5$ &  36.5 \\
   6145.02 &  5.61 & $-$1.42 & 1.3 & 9.8\,10$^5$ &  39.9 \\
   6155.14 &  5.62 & $-$0.75 & 1.3 & 3.6\,10$^6$ &  88.7 \\
   6848.57 &  5.86 & $-$1.65 & 1.3 & 1.1\,10$^6$ &  18.3 \\
   7455.39 &  5.96 & $-$2.01 & 1.3 & 4.0\,10$^5$ &   7.5 \\
   7800.00 &  6.18 & $-$0.73 & 1.3 & 3.3\,10$^6$ &  55.7 \\
\multicolumn{4}{l}{{\bf S I}~~~ $\log\epsilon_\odot=7.33$} \\
   6046.02 &  7.87 & $-$0.41 & 2.5 & 1.0\,10$^7$ &  17.3 \\
   6743.58 &  7.87 & $-$0.64 & 2.5 & 1.1\,10$^7$ &  10.3 \\
   7686.13 &  7.87 & $-$1.15 & 2.5 & 1.7\,10$^6$ &   3.4 \\
\multicolumn{4}{l}{{\bf K I}~~~ $\log\epsilon_\odot=5.12$} \\
   5801.75 &  1.62 & $-$1.51 & 2.5 & 3.1\,10$^6$ &   2.0 \\
   7698.98 &  0.00 &    0.01 &  A  & 5.8\,10$^7$ & 159.5 \\
\noalign{\smallskip}
\hline
\end{tabular}
\end{flushleft}
\end{table}

\setcounter{table}{1}
\begin{table}
\caption[]{
Continued
}
\begin{flushleft}
\begin{tabular}{rrrcrr}
\hline\noalign{\smallskip}
 $\lambda$~~~~ & $\chi_{\rm l}$~ & $\log gf$ & $\delta\Gamma_6$ & $\Gamma_{\rm rad}$~~~ & W$_{\lambda\odot}$ \\
 $[$\AA]~~~    & [eV]            &           &                  &       s$^{-1}$~~~     & [m\AA ]            \\
\noalign{\smallskip}
\hline\noalign{\smallskip}
\multicolumn{4}{l}{{\bf Ca I}~~~ $\log\epsilon_\odot=6.36$} \\
   5867.57 &  2.93 & $-$1.64 & 1.8 & 2.6\,10$^8$ &  25.3 \\
   6166.44 &  2.52 & $-$1.22 &  B  & 1.9\,10$^7$ &  71.7 \\
   6169.04 &  2.52 & $-$0.88 &  B  & 2.0\,10$^7$ &  96.0 \\
   6455.61 &  2.52 & $-$1.43 &  A  & 4.6\,10$^7$ &  58.6 \\
   6464.68 &  2.52 & $-$2.41 &  A  & 4.6\,10$^7$ &  12.6 \\
   6572.80 &  0.00 & $-$4.33 &  A  & 2.6\,10$^7$ &  33.6 \\
\multicolumn{4}{l}{{\bf Sc I}~~~ $\log\epsilon_\odot=3.17$} \\
   5520.51 &  1.87 &    0.62 &  A  & 9.3\,10$^7$ &   9.0 \\
\multicolumn{4}{l}{{\bf Sc II}~~~ $\log\epsilon_\odot=3.17$} \\
   5318.36 &  1.36 & $-$1.81 & 2.5 & 1.4\,10$^8$ &  13.5 \\
   6245.62 &  1.51 & $-$1.10 & 2.5 & 4.3\,10$^8$ &  36.7 \\
   6604.60 &  1.36 & $-$1.25 & 2.5 & 1.5\,10$^8$ &  37.0 \\
\multicolumn{4}{l}{{\bf Ti I}~~~ $\log\epsilon_\odot=5.02$} \\
   5113.45 &  1.44 & $-$0.91 &  A  & 2.2\,10$^7$ &  27.5 \\
   5219.71 &  0.02 & $-$2.29 &  A  & 6.7\,10$^6$ &  28.7 \\
   5426.26 &  0.02 & $-$3.11 &  A  & 1.7\,10$^6$ &   6.2 \\
   5766.33 &  3.29 &    0.26 &  B  & 1.4\,10$^8$ &   9.4 \\
   5866.46 &  1.07 & $-$0.88 &  A  & 6.4\,10$^7$ &  48.7 \\
   6091.18 &  2.27 & $-$0.47 &  A  & 8.5\,10$^7$ &  15.9 \\
   6126.22 &  1.07 & $-$1.44 &  A  & 9.9\,10$^6$ &  22.8 \\
   6258.11 &  1.44 & $-$0.47 &  A  & 1.7\,10$^8$ &  52.2 \\
\multicolumn{4}{l}{{\bf V I}~~~ $\log\epsilon_\odot=4.00$} \\
   5727.06 &  1.08 & $-$0.03 &  A  & 7.1\,10$^7$ &  39.4 \\
   6039.74 &  1.06 & $-$0.73 &  A  & 4.0\,10$^7$ &  13.3 \\
   6090.22 &  1.08 & $-$0.15 &  A  & 4.0\,10$^7$ &  34.2 \\
   6111.65 &  1.04 & $-$0.80 &  A  & 3.9\,10$^7$ &  12.1 \\
   6216.36 &  0.28 & $-$0.87 &  A  & 3.1\,10$^6$ &  38.4 \\
   6224.51 &  0.29 & $-$1.91 &  A  & 1.2\,10$^6$ &   5.7 \\
   6251.83 &  0.28 & $-$1.42 &  A  & 3.1\,10$^6$ &  15.8 \\
\multicolumn{4}{l}{{\bf Cr I}~~~ $\log\epsilon_\odot=5.67$} \\
   6330.10 &  0.94 & $-$2.94 &  A  & 2.4\,10$^7$ &  27.1 \\
\multicolumn{4}{l}{{\bf Cr II}~~~ $\log\epsilon_\odot=5.67$} \\
   5305.87 &  3.83 & $-$2.06 & 2.5 & 2.6\,10$^8$ &  26.6 \\
\multicolumn{4}{l}{{\bf Fe I}~~~ $\log\epsilon_\odot=7.50$} \\
   5067.16 &  4.22 & $-$0.96 &  B  & 2.1\,10$^8$ &  73.5 \\
   5090.78 &  4.26 & $-$0.61 &  B  & 2.1\,10$^8$ &  96.5 \\
   5109.66 &  4.30 & $-$0.82 &  B  & 2.1\,10$^8$ &  79.4 \\
   5141.75 &  2.42 & $-$2.20 &  A  & 1.5\,10$^8$ &  89.6 \\
   5358.12 &  3.30 & $-$3.20 &  A  & 2.0\,10$^8$ &  10.3 \\
   5809.22 &  3.88 & $-$1.70 &  A  & 5.1\,10$^7$ &  50.6 \\
   5849.69 &  3.69 & $-$2.98 &  A  & 5.5\,10$^7$ &   7.8 \\
   5852.23 &  4.55 & $-$1.23 &  B  & 1.9\,10$^8$ &  41.2 \\
   5855.09 &  4.61 & $-$1.57 &  B  & 1.9\,10$^8$ &  22.4 \\
   5856.10 &  4.29 & $-$1.60 & 1.4 & 8.6\,10$^7$ &  33.4 \\
   5858.79 &  4.22 & $-$2.20 &  B  & 2.7\,10$^8$ &  13.8 \\
   5859.60 &  4.55 & $-$0.71 &  B  & 1.9\,10$^8$ &  72.8 \\
   5861.11 &  4.28 & $-$2.36 &  B  & 2.1\,10$^8$ &   8.9 \\
   5862.37 &  4.55 & $-$0.49 &  B  & 1.9\,10$^8$ &  88.4 \\
   6151.62 &  2.18 & $-$3.30 &  A  & 1.6\,10$^8$ &  50.9 \\
   6157.73 &  4.08 & $-$1.22 & 1.4 & 5.0\,10$^7$ &  63.4 \\
   6159.38 &  4.61 & $-$1.88 &  B  & 1.9\,10$^8$ &  12.6 \\
   6165.36 &  4.14 & $-$1.50 & 1.4 & 8.8\,10$^7$ &  46.3 \\
   6173.34 &  2.22 & $-$2.88 &  A  & 1.7\,10$^8$ &  69.7 \\
   6200.32 &  2.61 & $-$2.39 &  A  & 1.0\,10$^8$ &  75.3 \\
   6436.41 &  4.19 & $-$2.39 & 1.4 & 3.0\,10$^7$ &  10.2 \\
\noalign{\smallskip}
\hline
\end{tabular}
\end{flushleft}
\end{table}

\setcounter{table}{1}
\begin{table}
\caption[]{
Continued
}
\begin{flushleft}
\begin{tabular}{rrrcrr}
\hline\noalign{\smallskip}
 $\lambda$~~~~ & $\chi_{\rm l}$~ & $\log gf$ & $\delta\Gamma_6$ & $\Gamma_{\rm rad}$~~~ & W$_{\lambda\odot}$ \\
 $[$\AA]~~~    & [eV]            &           &                  &       s$^{-1}$~~~     & [m\AA ]            \\
\noalign{\smallskip}
\hline\noalign{\smallskip}
\multicolumn{4}{l}{{\bf Fe I} cont.~~~ $\log\epsilon_\odot=7.50$} \\
   6591.33 &  4.59 & $-$1.99 & 1.4 & 1.4\,10$^8$ &  10.8 \\
   6608.04 &  2.28 & $-$3.95 &  A  & 1.7\,10$^8$ &  18.7 \\
   6699.14 &  4.59 & $-$2.12 & 1.4 & 1.4\,10$^8$ &   8.4 \\
   6713.75 &  4.80 & $-$1.44 &  B  & 2.4\,10$^8$ &  21.5 \\
   6725.36 &  4.10 & $-$2.21 &  A  & 2.1\,10$^8$ &  17.9 \\
   6732.07 &  4.58 & $-$2.21 & 1.4 & 6.0\,10$^7$ &   7.1 \\
   6733.15 &  4.64 & $-$1.43 & 1.4 & 2.3\,10$^8$ &  27.7 \\
   6857.25 &  4.08 & $-$2.07 & 1.4 & 2.5\,10$^7$ &  23.5 \\
   7751.12 &  4.99 & $-$0.80 &  B  & 6.4\,10$^8$ &  46.5 \\
   7802.51 &  5.09 & $-$1.36 &  B  & 6.3\,10$^8$ &  16.1 \\
   7844.57 &  4.84 & $-$1.71 & 1.4 & 2.3\,10$^8$ &  12.8 \\
   8747.44 &  3.02 & $-$3.35 &  A  & 8.0\,10$^7$ &  18.1 \\
   8757.20 &  2.84 & $-$2.01 &  A  & 7.7\,10$^6$ &  97.4 \\
\multicolumn{4}{l}{{\bf Fe II}~~~ $\log\epsilon_\odot=7.50$} \\
   5100.66 &  2.81 & $-$4.15 & 2.5 & 3.4\,10$^8$ &  22.1 \\
   5256.93 &  2.89 & $-$4.08 & 2.5 & 3.4\,10$^8$ &  22.1 \\
   5425.26 &  3.20 & $-$3.31 & 2.5 & 3.0\,10$^8$ &  41.8 \\
   5427.80 &  6.72 & $-$1.53 & 2.5 & 3.5\,10$^8$ &   4.4 \\
   6149.25 &  3.89 & $-$2.78 & 2.5 & 3.4\,10$^8$ &  37.0 \\
   6247.56 &  3.89 & $-$2.41 & 2.5 & 3.4\,10$^8$ &  54.4 \\
   6369.46 &  2.89 & $-$4.16 & 2.5 & 2.9\,10$^8$ &  19.9 \\
   6383.72 &  5.55 & $-$2.11 & 2.5 & 4.1\,10$^8$ &   9.8 \\
   6432.68 &  2.89 & $-$3.64 & 2.5 & 2.9\,10$^8$ &  41.6 \\
   6456.39 &  3.90 & $-$2.20 & 2.5 & 3.4\,10$^8$ &  64.7 \\
   7479.70 &  3.89 & $-$3.67 & 2.5 & 3.1\,10$^8$ &   9.3 \\
   7515.84 &  3.90 & $-$3.50 & 2.5 & 4.1\,10$^8$ &  12.6 \\
   7841.37 &  3.90 & $-$4.04 & 2.5 & 3.0\,10$^8$ &   4.3 \\
\multicolumn{4}{l}{{\bf Ni I}~~~ $\log\epsilon_\odot=6.25$} \\
   5082.35 &  3.66 & $-$0.59 &  B  & 1.8\,10$^8$ &  67.9 \\
   5084.11 &  3.68 & $-$0.13 &  B  & 1.3\,10$^8$ &  94.8 \\
   5088.54 &  3.85 & $-$1.08 &  B  & 1.6\,10$^8$ &  33.5 \\
   5088.96 &  3.68 & $-$1.30 &  B  & 1.3\,10$^8$ &  30.4 \\
   5094.42 &  3.83 & $-$1.14 &  B  & 1.2\,10$^8$ &  31.3 \\
   5102.97 &  1.68 & $-$2.77 &  A  & 1.3\,10$^8$ &  51.3 \\
   5115.40 &  3.83 & $-$0.35 & 2.5 & 7.3\,10$^7$ &  78.6 \\
   5847.01 &  1.68 & $-$3.44 &  A  & 7.4\,10$^7$ &  23.3 \\
   6111.08 &  4.09 & $-$0.83 &  B  & 1.5\,10$^8$ &  36.1 \\
   6130.14 &  4.27 & $-$0.97 &  B  & 2.8\,10$^8$ &  22.4 \\
   6133.98 &  4.09 & $-$1.80 &  B  & 1.4\,10$^8$ &   6.1 \\
   6175.37 &  4.09 & $-$0.55 &  B  & 2.3\,10$^8$ &  50.9 \\
   6176.82 &  4.09 & $-$0.29 &  B  & 1.5\,10$^8$ &  65.4 \\
   6177.25 &  1.83 & $-$3.54 & 2.5 & 4.3\,10$^7$ &  15.3 \\
   6204.61 &  4.09 & $-$1.14 &  A  & 1.8\,10$^8$ &  23.2 \\
   6378.26 &  4.15 & $-$0.85 &  B  & 2.1\,10$^8$ &  32.8 \\
   6850.44 &  3.68 & $-$1.99 &  A  & 9.9\,10$^7$ &  10.0 \\
   7748.89 &  3.71 & $-$0.39 &  A  & 9.5\,10$^7$ &  90.5 \\
   7797.59 &  3.90 & $-$0.38 &  A  & 1.0\,10$^8$ &  80.6 \\
   7826.77 &  3.70 & $-$1.87 &  A  & 8.0\,10$^7$ &  13.1 \\
\multicolumn{4}{l}{{\bf Y II}~~~ $\log\epsilon_\odot=2.24$} \\
   5087.43 &  1.08 & $-$0.36 & 2.5 & 1.3\,10$^7$ &  48.3 \\
   5200.42 &  0.99 & $-$0.65 & 2.5 & 1.1\,10$^7$ &  39.9 \\
   5402.78 &  1.84 & $-$0.58 & 2.5 & 8.6\,10$^6$ &  12.8 \\
\multicolumn{4}{l}{{\bf Zr I}~~~ $\log\epsilon_\odot=2.60$} \\
   6127.48 &  0.15 & $-$0.92 &  A  & 2.4\,10$^6$ &   3.8 \\
   6134.57 &  0.00 & $-$1.20 &  A  & 2.2\,10$^6$ &   2.8 \\
\noalign{\smallskip}
\hline
\end{tabular}
\end{flushleft}
\end{table}

\setcounter{table}{1}
\begin{table}
\caption[]{
Continued
}
\begin{flushleft}
\begin{tabular}{rrrcrr}
\hline\noalign{\smallskip}
 $\lambda$~~~~ & $\chi_{\rm l}$~ & $\log gf$ & $\delta\Gamma_6$ & $\Gamma_{\rm rad}$~~~ & W$_{\lambda\odot}$ \\
 $[$\AA]~~~    & [eV]            &           &                  &       s$^{-1}$~~~     & [m\AA ]            \\
\noalign{\smallskip}
\hline\noalign{\smallskip}
\multicolumn{4}{l}{{\bf Zr II}~~~ $\log\epsilon_\odot=2.60$} \\
   5112.28 &  1.66 & $-$0.78 & 2.5 & 1.1\,10$^7$ &   9.7 \\
\multicolumn{4}{l}{{\bf Ba II}~~~ $\log\epsilon_\odot=2.13$} \\
   5853.69 &  0.60 & $-$0.99 & 3.0 & 1.6\,10$^8$ &  64.2 \\
   6141.73 &  0.70 & $-$0.08 & 3.0 & 1.6\,10$^8$ & 119.5 \\
   6496.91 &  0.60 & $-$0.40 & 3.0 & 1.3\,10$^8$ & 102.7 \\
\multicolumn{4}{l}{{\bf Eu II}~~~ $\log\epsilon_\odot=0.51$} \\
   6645.13 &  1.38 &    0.24 & 2.5 & 2.4\,10$^7$ &   5.3 \\
\noalign{\smallskip}
\hline
\end{tabular}
\end{flushleft}
\end{table}

A few lines were excluded from the analysis after a first abundance analysis of
all stars, since they give abundances that deviate strongly, often as a
function of effective temperature, from those of other lines of the same
species:
Sc\,{\sc i} 6239.36, Ti\,{\sc i} 5299.98, 
V\,{\sc i} 6150.15, Fe\,{\sc i} 5127.37, Ni\,{\sc i} 5099.94 and
7788.15 and Y\,{\sc i} 6687.51\,\AA.
For some of these, asymmetries could be traced in the spectra, while for others
we suspect undetected blends or line misidentifications.
For the 7788.15\,\AA\ Ni\,{\sc i} line a laboratory oscillator strength is quoted
in the VALD data base which is 0.5\,dex lower than our solar value, which means
that most of the observed line absorption in the Sun may be caused by another
transition.  Therefore, these lines were not used.
Also the Sc\,{\sc ii} 5318.36\,\AA\ line was found to be asymmetric in the two
hottest stars and was not used for 9~Com and HR~8472.

The solar equivalent widths and adopted line data are given in Table~2.

   \subsection{Model atmosphere parameters}

Initially, the model parameters $T_{\rm eff}$, $\log g$ and [Fe/H] were adopted
from EAGLNT, as well as microturbulence parameters computed from their Eq. (9).
The larger set of observed spectral lines in this project enables the
application of better spectroscopic statistical checks on the individual stellar
effective temperatures, microturbulence parameters and surface gravities.
The initial values were then iteratively modified as described below.
Since, for metal-rich stars, the temperature structures of the model atmospheres
are rather sensitive to the overall metallicity, the model metallicities were
kept consistent with the [Fe/H] values obtained in the analysis.
\begin{figure}
\centerline{\psfig{figure=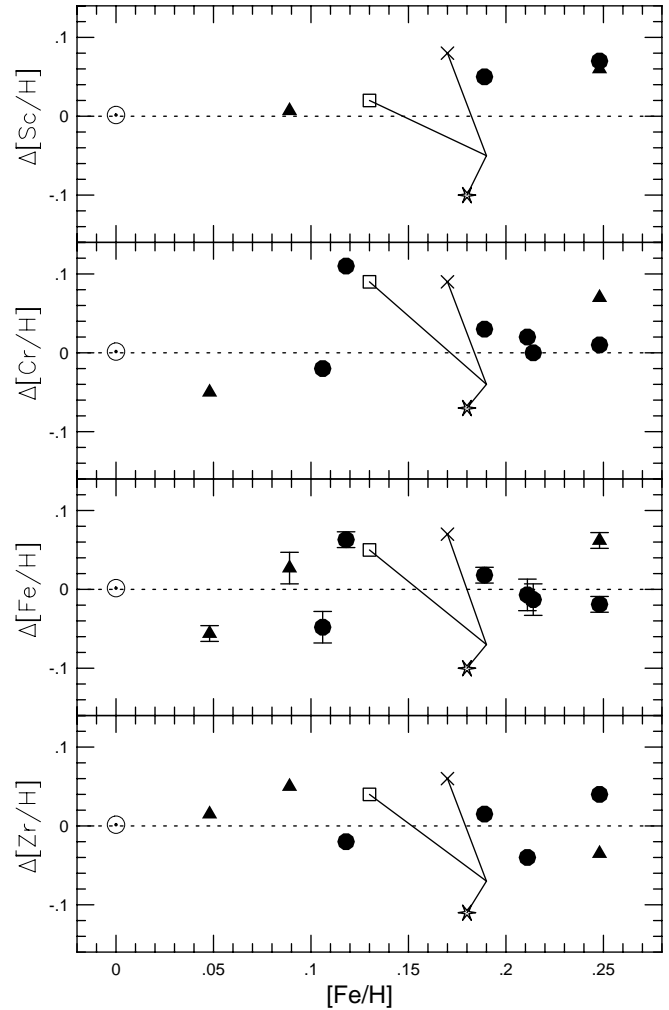,width=9cm}}
\caption[]{
Comparison of abundances derived from lines of the neutral and singly ionized
species of the same elements, e.g., 
$\Delta$[Sc/H]~= [Sc/H](lines~of~Sc\,{\sc ii})~$-$
                 [Sc/H](lines~of~Sc\,{\sc i}).
Stars classified as NaMgAl stars are shown as filled circles, and non-NaMgAl
stars as filled triangles.
The typical effects on the ratios of a lowering of the effective temperature
by 100\,K are shown by the lines which end with a square symbol.
Likewise the effects of an increase of $\log g$ by 0.3\,dex is shown by the
crosses and of that a lowering of the model metallicity by 0.2\,dex by the
stars
}
\end{figure}

\begin{enumerate}

\item Erroneous effective temperatures show up (assuming) LTE
as systematic dependences of derived line-by-line abundances as a function of
line excitation energy.
To investigate such effects, the abundances of Fe\,{\sc i} and Ni\,{\sc i}
lines with reduced
equivalent widths less than $\log W_\lambda({\rm theor})/\lambda=-5.1$
(to reduce the influence of possible uncertainties in the line broadening)
were plotted versus the excitation energy of the lower level.
To avoid systematic errors due to random equivalent width measurement errors,
theoretical equivalent widths, $W_\lambda({\rm theor})$, as suggested by
Magain (1984) were used for this check.
The effective temperature of each star was then modified to minimise these
slopes.

\item Similarly, inconsistent microturbulences should result in
abundances which vary as a function of line strength for different lines of a
species.
For this test the Fe\,{\sc i} and Ni\,{\sc i} lines with excitation energies
$\ge 4.0$\,eV (to minimise possible effects of errors in the
excitation balances) were plotted against $\log W_\lambda({\rm theor})/\lambda$.
The microturbulence parameter of each star was then modified to minimise these
slopes.

\item The surface gravities were then determined by synthesis of the spectral
region near the strong Ca\,{\sc i} 6162.17\,\AA\ line, the wings of which are
sensitive to the gas pressure and therefore to surface gravity
(cf. e.g. Edvardsson 1988).

\item As a fourth consistency check, the ionization balances, which in the LTE
assumption are sensitive to the effective temperatures and surface gravities via
the Saha equation, were studied for four elements which are represented by
both neutral and singly ionised lines in the investigation: Sc, Cr, Fe and Zr.
While there are many Fe\,{\sc i} and Fe\,{\sc ii} lines measured, the numbers
of measured lines for the other species are few: only one line
for Sc\,{\sc i}, Cr\,{\sc i}, Cr\,{\sc ii} and Zr\,{\sc ii}, and at most three
lines for Sc\,{\sc ii} and Zr\,{\sc i}.
Nevertheless, all the four elements show very similar sensitivities to the
stellar parameters and give consistent indications of the surface gravity for
a reasonable choice of effective temperature.
This is demonstrated in Fig.~3.

\end{enumerate}

These four checks were then performed while the atmospheric parameters were
iterated until good consistency was obtained for each individual star.
The finally adopted atmospheric parameters are given in Table~1.

It is now of interest to investigate how these parameters deviate from the
values in EAGLNT, which were determined from Str\"omgren photometry.
The new effective temperatures are $45$\,K with a $1\,\sigma$ dispersionof
$\pm 35$\,K hotter, which is actually
somewhat less than what was found in an excitation-equilibrium test
in Sect. 4.3.4 of EAGLNT.
The new surface gravities are $0.17$\,dex with a dispersion of $\pm 0.13$\,dex
higher than those of EAGLNT,
which is within the total expected uncertainty given in their Sect. 4.2.1.
Our [Fe/H] values differ from the photometric metallicities of EAGLNT by
$0.00 \pm 0.12$\,dex (dispersion), and finally the microturbulence parameters
are
$+0.05 \pm 0.09$\,km\,s$^{-1}$ higher than Eq. 9 of EAGLNT would give for the
new values of $T_{\rm eff}$ and $\log g$.
No systematic dependence of these differences on the atmospheric parameters
themselves can be seen.

   \subsection{Errors in the resulting abundances}

An extensive discussion of error sources (e.g., NLTE-effects and
oversimplification of the model atmospheres) in an analysis which is very
similar to the present one was given by EAGLNT, and will not be repeated here.
We note, however, that we now have a much larger sample of spectral lines, and
that the equivalent widths are more accurate.
Furthermore, the present sample of stars spans a small range in metallicity
and a range of only 10\% in effective temperature, so much of the actual
errors cancel in an analysis relative to the Sun.

Figure~3 shows the remaining small differences between the absolute abundances
of an element as derived from two different states of ionization.
Only for iron are there several lines measured from both the neutral and the
ionized species, and the error bars show the formal errors of the mean
abundances (calculated from the internal scatter among Fe\,{\sc i} and
Fe\,{\sc ii} lines, respectively) added in quadrature.

Also the abundance ratios relative to Fe, [X/Fe], for these elements are very
similar if data from neutral and ionized lines are used, as can be seen in
Fig.~4.
This confirms the good quality of the equivalent width measurements.
\begin{figure}
\centerline{\psfig{figure=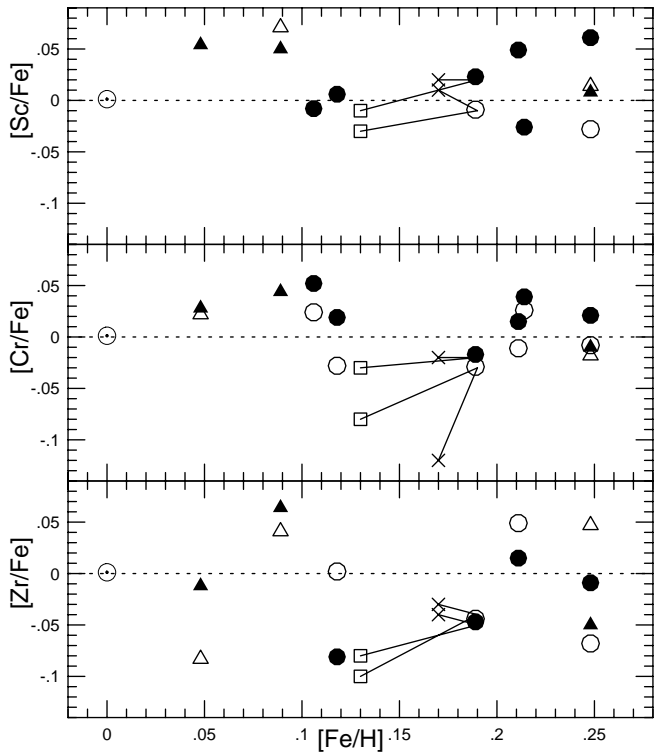,width=9cm}}
\caption[]{
Relative abundances [X/Fe] (X = Sc, Cr and Zr) where the open symbols represent
results from lines of X\,{\sc i} and Fe\,{\sc i} and the filled symbols show
ratios derived from X\,{\sc ii} and Fe\,{\sc ii} lines.
Circles represent NaMgAl stars and triangles non-NaMgAl stars.
The small differences between abundance ratios derived from neutral and ionized
lines testify to the accuracy of the measured equivalent widths.
The effects of changes to $T_{\rm eff}$ and $\log g$ are indicated as in
Fig.~3
}
\end{figure}

For C, N, O and S, the (high-excitation) lines used are of the neutral atom
which is the dominant species due to their high ionization energies.
When abundance ratios to the iron atom are plotted one finds that the ratios
decrease systematically with increasing metallicity.
If the ratios are instead taken with respect to the majority species
Fe\,{\sc ii} the amounts of scatter around the trends diminish appreciably.
This is shown in Fig.~5.
The sensitivity of the abundance ratios to effective temperature and surface
gravity are clearly different, which, together with the smaller scatter,
suggests that the ratios relative to Fe\,{\sc ii} are the more accurate.
\begin{figure*}
\centerline{\psfig{figure=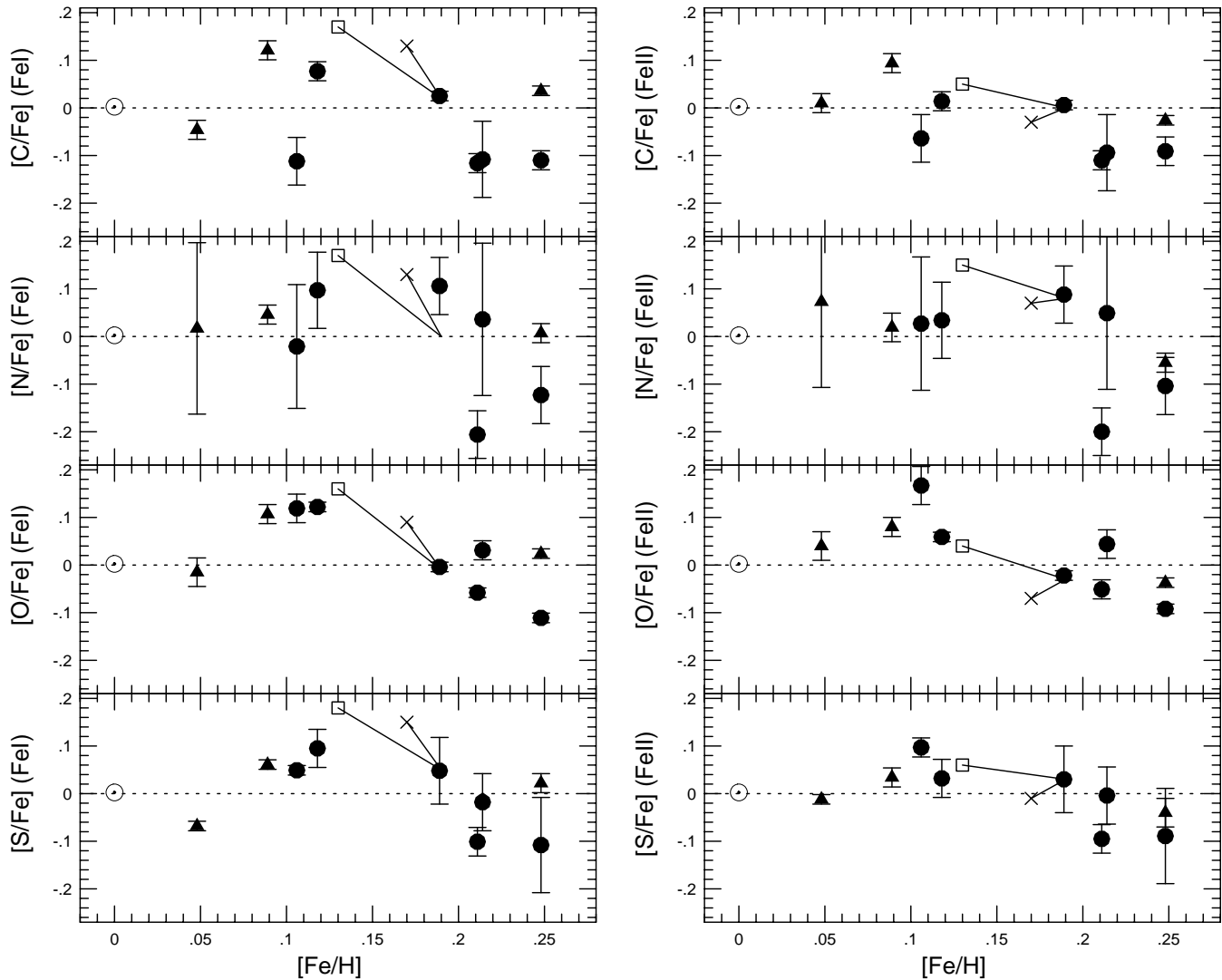,width=18cm}}
\caption[]{
Abundance ratios of C, N, O and S (derived from high excitation lines of the
neutral atoms) plotted vs. iron abundance.
The left hand panels show ratios relative to Fe\,{\sc i} lines, and the right
hand panels show ratios relative to Fe\,{\sc ii} lines.
The standard deviations of the abundances due to the scatter among the results
for individual lines are shown by the error bars.
The effects of lowering of the effective temperatures by 100\,K are shown for
one star by the open squares, and those of a doubling of the surface gravity
by the crosses
}
\end{figure*}


   \section{Results}

The resulting abundances for the nine stars are given in Table~3, and the
line-by-line measured equivalent widths and abundances are given in Table~4
(available in electronic form).
\begin{table*}  
\caption[]{
Abundances relative to hydrogen [X/H] derived for the programme stars.
$N$ is the number of lines used in the analysis for a species.
The quoted errors are only the standard deviations in the mean values due to
the line-to-line scatter within the species with an absent error figure
indicating that only one line was measured
}
\begin{flushleft}  
\begin{tabular}{lclllllllll} 
\noalign{\smallskip}
\hline
\noalign{\smallskip}
Species & $N$ & HR~448 & HR~1536 & HR~3176 & HR~3951 & HR~4027 & HR~4688 & HR~8041 & HR~8472 & HR~8729 \\
        &   &    &    & $\mu$~Cnc~~ & 20~LMi~ & 24~LMi~ & 9~Com~ & 11~Aqr~ &  & 51~Peg~ \\
\hline
Fe\,{\sc  i} & 22--34 & 0.12$\pm$.01 & 0.21$\pm$.01 & 0.09$\pm$.01 & 0.25$\pm$.00 & 0.05$\pm$.00 & 0.21$\pm$.01 & 0.25$\pm$.00 & 0.11$\pm$.01 & 0.19$\pm$.00  \\
 C\,{\sc  i} &  3--6  & 0.20$\pm$.02 & 0.09$\pm$.02 & 0.21$\pm$.02 & 0.14$\pm$.02 & 0.00$\pm$.02 & 0.11$\pm$.08 & 0.28$\pm$.01 &-0.01$\pm$.05 & 0.21$\pm$.01  \\
 N\,{\sc  i} &  2     & 0.22$\pm$.08 & 0.00$\pm$.05 & 0.14$\pm$.02 & 0.12$\pm$.06 & 0.06$\pm$.18 & 0.25$\pm$.16 & 0.25$\pm$.02 & 0.09$\pm$.13 & 0.30$\pm$.06  \\
 O\,{\sc  i} &  3--4  & 0.24$\pm$.01 & 0.15$\pm$.00 & 0.20$\pm$.01 & 0.14$\pm$.00 & 0.03$\pm$.03 & 0.25$\pm$.02 & 0.27$\pm$.01 & 0.22$\pm$.03 & 0.19$\pm$.00  \\
Na\,{\sc  i} &  2--3  & 0.22$\pm$.01 & 0.23$\pm$.01 & 0.20$\pm$.01 & 0.27$\pm$.01 & 0.06$\pm$.00 & 0.30$\pm$.01 & 0.35$\pm$.01 & 0.17$\pm$.02 & 0.26$\pm$.01  \\
Mg\,{\sc  i} &  1--4  & 0.23$\pm$.06 & 0.21$\pm$.05 & 0.14$\pm$.05 & 0.23$\pm$.05 & 0.06$\pm$.06 & 0.07         & 0.23$\pm$.05 & 0.08$\pm$.16 & 0.18$\pm$.10  \\
Al\,{\sc  i} &  4--7  & 0.19$\pm$.01 & 0.26$\pm$.01 & 0.19$\pm$.01 & 0.27$\pm$.01 & 0.10$\pm$.01 & 0.26$\pm$.02 & 0.27$\pm$.01 & 0.16$\pm$.01 & 0.21$\pm$.01  \\
Si\,{\sc  i} &  5--7  & 0.20$\pm$.01 & 0.25$\pm$.01 & 0.15$\pm$.01 & 0.26$\pm$.01 & 0.07$\pm$.01 & 0.25$\pm$.01 & 0.30$\pm$.01 & 0.17$\pm$.02 & 0.24$\pm$.01  \\
 S\,{\sc  i} &  2--3  & 0.21$\pm$.03 & 0.11$\pm$.03 & 0.15$\pm$.01 & 0.14$\pm$.10 &-0.02$\pm$.01 & 0.20$\pm$.05 & 0.27$\pm$.02 & 0.16$\pm$.01 & 0.24$\pm$.07  \\
 K\,{\sc  i} &  0--2  & 0.16         & 0.21         & 0.13         & 0.16$\pm$.00 & 0.16$\pm$.04 &  ---         & 0.18$\pm$.03 & 0.27         & 0.15$\pm$.01  \\
Ca\,{\sc  i} &  3--6  & 0.10$\pm$.01 & 0.23$\pm$.01 & 0.07$\pm$.01 & 0.25$\pm$.01 & 0.06$\pm$.01 & 0.20$\pm$.01 & 0.20$\pm$.01 & 0.10$\pm$.04 & 0.16$\pm$.00  \\
Sc\,{\sc  i} &  0--1  &  ---         &  ---         & 0.16         & 0.22         &  ---         &  ---         & 0.26         &  ---         & 0.18          \\
Sc\,{\sc ii} &  2--3  & 0.19$\pm$.02 & 0.25$\pm$.02 & 0.17$\pm$.04 & 0.29$\pm$.02 & 0.05$\pm$.03 & 0.17$\pm$.02 & 0.32$\pm$.02 & 0.05$\pm$.03 & 0.23$\pm$.01  \\
Ti\,{\sc  i} &  4--8  & 0.11$\pm$.01 & 0.25$\pm$.01 & 0.11$\pm$.02 & 0.28$\pm$.01 & 0.09$\pm$.02 & 0.24$\pm$.03 & 0.23$\pm$.01 & 0.10$\pm$.03 & 0.19$\pm$.01  \\
 V\,{\sc  i} &  4--7  & 0.11$\pm$.01 & 0.26$\pm$.01 & 0.11$\pm$.02 & 0.31$\pm$.01 & 0.07$\pm$.02 & 0.21$\pm$.02 & 0.27$\pm$.01 & 0.08$\pm$.02 & 0.22$\pm$.01  \\
Cr\,{\sc  i} &  0--1  & 0.09         & 0.20         &  ---         & 0.24         & 0.07         & 0.24         & 0.23         & 0.13         & 0.16          \\
Cr\,{\sc ii} &  1     & 0.20         & 0.22         & 0.16         & 0.25         & 0.02         & 0.24         & 0.30         & 0.11         & 0.19          \\
Fe\,{\sc ii} &  8--13 & 0.18$\pm$.01 & 0.20$\pm$.02 & 0.12$\pm$.01 & 0.23$\pm$.01 &-0.01$\pm$.01 & 0.20$\pm$.02 & 0.31$\pm$.01 & 0.06$\pm$.02 & 0.21$\pm$.01  \\
Ni\,{\sc  i} & 11--20 & 0.15$\pm$.01 & 0.25$\pm$.01 & 0.13$\pm$.01 & 0.27$\pm$.01 & 0.04$\pm$.00 & 0.23$\pm$.01 & 0.30$\pm$.01 & 0.11$\pm$.02 & 0.23$\pm$.00  \\
 Y\,{\sc ii} &  3--3  & 0.08$\pm$.03 & 0.21$\pm$.01 & 0.06$\pm$.03 & 0.24$\pm$.01 &-0.07$\pm$.02 & 0.20$\pm$.02 & 0.22$\pm$.01 & 0.09$\pm$.04 & 0.15$\pm$.01  \\
Zr\,{\sc  i} &  0--2  & 0.12$\pm$.10 & 0.26$\pm$.11 & 0.13         & 0.18$\pm$.00 &-0.04$\pm$.06 &  ---         & 0.30$\pm$.05 &  ---         & 0.15$\pm$.08  \\
Zr\,{\sc ii} &  0--1  & 0.10         & 0.22         & 0.18         & 0.22         &-0.02         &  ---         & 0.26         &  ---         & 0.16          \\
Ba\,{\sc ii} &  2--3  & 0.04$\pm$.02 & 0.17$\pm$.01 &-0.03$\pm$.01 & 0.21$\pm$.01 &-0.03$\pm$.01 & 0.12$\pm$.01 & 0.16$\pm$.01 & 0.04$\pm$.04 & 0.12$\pm$.01  \\
\hline
\end{tabular}
\end{flushleft}
\end{table*}

Most of the elements under investigation have ionization
energies and partition functions such that the majority of the atoms are in the
singly ionized state in the line-forming layers of these solar-type stars.
When we display abundance ratios between different elements, we will prefer
to display ratios using abundances derived from lines of the same ionization
state in the nominator and the denominator.
This procedure minimises the effects of remaining uncertainties in the model
atmospheres, in the assumptions made in the analysis and in
the stellar fundamental parameters.
The exceptions to this rule are C, N, O and S as discussed above.

Fig.~6 displays our abundance results, where abundance ratios are plotted as a
function of [Fe/H] derived from Fe\,{\sc I} lines.
In the figure we have also indicated the sensitivity of the abundance ratios to
uncertainties in the effective temperatures and surface gravities.
The results for C, N, O and S were already shown in Fig.~5.
\begin{figure*}
\centerline{\psfig{figure=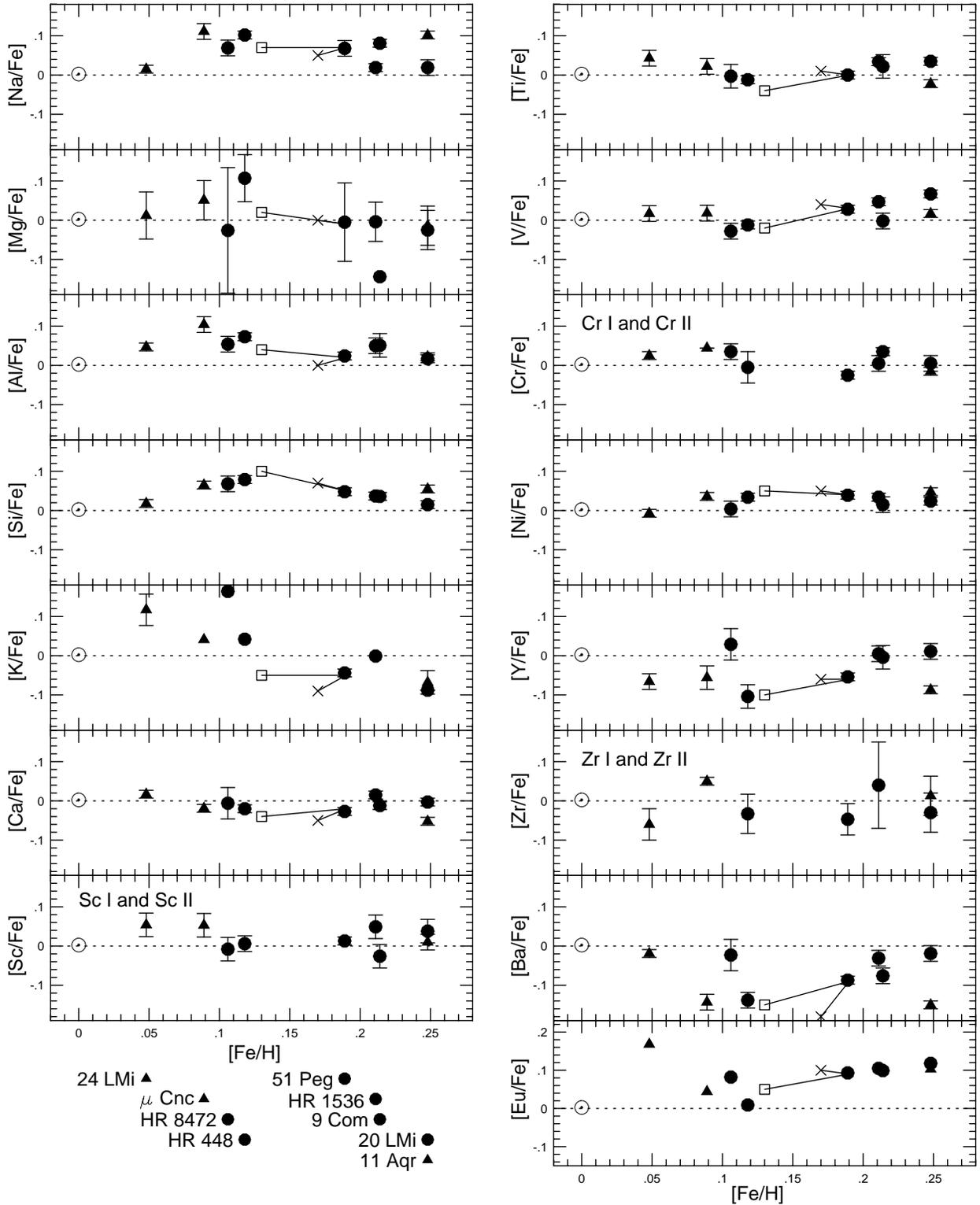,width=17cm}}
\caption[]{
Abundance ratios between different elements as a function of the iron
abundance derived from Fe\,{\sc i} lines.
The error bars show the statistical standard deviation of the mean of the
abundance ratio derived from the scatter among the individual lines.
For Sc, Cr and Zr, the plots show the mean of the abundances derived from
neutral and ionized lines weighted by the available number of lines.
For 51~Peg, the squares show the shift caused by a lowering of
$T_{\rm eff}$ by 100\,K, and the crosses show the effect of increasing the
surface gravity by 0.3\,dex. For Sc, Cr and Zr, these effects are shown in
Fig.~4 for neutral and ionized lines separately.
The stars are identified by symbol and [Fe/H] in the lower left corner of the
figure
}
\end{figure*}

In spite of the small number of stars the high accuracy in the
analysis makes it worthwhile to compare the results with abundance trends 
found in other analyses.
We postpone, however, this discussion to Feltzing \& Gustafsson (1997).

Inspection of Fig.~6 shows there are no significant differences between
the [Na/Fe], [Mg/Fe] and [Al/Fe] ratios of the NaMgAl stars
and the non-NaMgAl stars.
The six NaMgAl stars have average abundances
[Na/Fe] $= 0.06 \pm 0.02$ (standard deviation of the mean),
[Mg/Fe] $= -0.02 \pm 0.04$ and [Al/Fe] $= 0.04 \pm 0.01$,
while the three stars not identified as NaMgAl stars have
[Na/Fe] $= 0.07 \pm 0.04$, [Mg/Fe] $= 0.01 \pm 0.02$ and
[Al/Fe] $= 0.06 \pm 0.03$.  (These results use abundances, including
those of Fe, from neutral lines.)  A larger number of stars not identified
as NaMgAl stars may be found in EAGLNT; for the 18 non-NaMgAl
stars of EAGLNT with [Fe/H] $\geq +0.05$ and abundances
derived from EAGLNT's high resolution (80\,000) ESO observations, we find
average abundances of [Na/Fe] $= 0.06 \pm 0.02$, [Mg/Fe] $= 0.07 \pm 0.01$
and [Al/Fe] $= 0.08 \pm 0.02$.  Evidently, the NaMgAl stars have no
significant enhancements of Na, Mg or Al and any possible enrichment
of these elements is limited to $\sim 0.03$\,dex, or less.

In view of our failure to confirm the Na, Mg and Al overabundances of
the NaMgAl stars claimed by EAGLNT, we now briefly review the sources
of difference between our new abundances and those of EAGLNT.  Five
primary potential sources of differences between the abundances of
the two investigations may be listed:
\begin{enumerate}

\item Use of additional/different lines

\item Different equivalent widths, stellar and solar

\item Different model parameters

\item Different types of models

\item Different $gf$ values

\end{enumerate}

The two investigations use the same types of model atmospheres -
line blanketed, flux constant, LTE models calculated from a
development of the programmes of Gustafsson et al. (1975) - and
both use astrophysical oscillator strengths derived from the
lines' equivalent widths in the solar flux spectrum.  Item four
is, thus, not a factor in the comparison, while item five is
subsumed under item two.  The causes of the differences between the
two investigations are, therefore, to be found in items one, two and three. 
(A comparison of [Fe/H] abundances with EAGLNT shows that the present [Fe/H]
values are significantly higher, by 0.08\,dex at a mean.
This is for most stars due to the systematically somewhat higher $T_{\rm eff}$s
and $\log g$s and lower microturbulence parameters of the present analysis, and
is within the range of systemic errors ascribed by EAGLNT to their [Fe/H]
values.
For 51~Peg the larger equivalent widths found here (Fig.~2) are the main cause.)
We now do a budget on a line-by-line basis for Na, Mg and Al
of the sources of the abundance differences for one of the stars
(51~Peg) which EAGLNT identified as a NaMgAl star.

In Table~5 we compare our new equivalent
widths and EAGLNT's for 51 Peg and the Sun for the
Na\,{\sc i} and Al\,{\sc i} lines in common to our investigation
and theirs; the two investigations do not have any Mg lines in common.
A comparison of our new model atmosphere parameters for 51~Peg
(Table~1) and EAGLNT's reveals modest revisions, namely an increase
of 20\,K in the effective temperature, an increase of 0.17 in
$\log g$, a decrease of 0.22\,km\,s$^{-1}$ in microturbulence and
an increase of 0.11\,dex in [M/H].  The separate effects on the
Na and Al abundances of the equivalent width and $\log gf$ changes
and these revisions of the model parameters are also shown in Table~5.
For the Na\,{\sc i} and Al\,{\sc i} lines in 
common to the two analyses the new analysis
gives [Na/H] = +0.25 and [Al/H] = +0.21, while for the same
lines EAGLNT found [Na/H] = +0.22 and [Al/H] = +0.28.  
The abundance differences, for the same lines, between the
new analysis and EAGLNT's, therefore, are $\Delta$[Na/H] = +0.03
and $\Delta$[Al/H] = --0.07, which for both elements differ
by only 0.01\,dex from the total of the separate effects
listed in Table~5.  For the
lines in common to the two investigations, therefore, the changes in the
Na and Al abundances are almost entirely due to changes of the stellar
equivalent widths, the $\log gf$ and the model atmosphere
parameters.
\setcounter{table}{4}
\begin{table*}  
\caption[]{
Comparison of new equivalent widths and EAGLNT's for Na\,{\sc i} and 
Al\,{\sc i} lines
}
\begin{flushleft}  
\begin{tabular}{rrrrrrrrrrrrrrr} 
\noalign{\smallskip}
\hline
\noalign{\smallskip}
Line~ & $\lambda$\,[\AA] & \multicolumn{2}{c}{W$_\lambda$\,[m\AA] new} &\ & \multicolumn{2}{c}{W$_\lambda$\,[m\AA] EAGLNT} &\ & 
\multicolumn{7}{c}{$\Delta$[X/H] (new -- EAGLNT)} \\
\cline{3-4}\cline{6-7}\cline{9-15}
   &      & 51 Peg & Sun &\ & 51 Peg & Sun &\ & W$_\lambda$ & $\log gf$ & $T_{\rm eff}$ & $\log g$ & $\xi_{\rm t}$ 
& [M/H] & Total \\
\hline
Na\,{\sc i}$^{\hspace{1ex}}$ & 6154.23 &  56.3 & 39.6 &\ &  54.1 & 38.2 &\ &  +0.03 & --0.02 & +0.01 & --0.01 & +0.02 & +0.01 & +0.04 \\
Na\,{\sc i}$^{\hspace{1ex}}$ & 6160.75 &  74.8 & 57.7 &\ &  74.9 & 58.6 &\ &   0.00 &  +0.01 & +0.01 & --0.01 & +0.02 & +0.01 & +0.04 \\
Average$^{\hspace{1ex}}$     &         &       &      &\ &       &      &\ &        &        &       &        &       &       & +0.04 \\
                                                                                    \\
Al\,{\sc i}$^a$              & 8772.88 &  93.8 & 75.2 &\ & 109.0 & 80.4 &\ & --0.15 &  +0.05 & +0.01 & --0.04 & +0.02 & +0.01 & --0.10 \\
Al\,{\sc i}$^{\hspace{1ex}}$ & 8773.91 & 114.1 & 94.4 &\ & 121.0 & 98.1 &\ & --0.06 &  +0.04 & +0.01 & --0.04 & +0.02 & +0.01 & --0.02 \\
Average$^{\hspace{1ex}}$     &         &       &      &\ &       &      &\ &        &        &       &        &       &       & --0.06 \\
\hline
\\
\multicolumn{15}{l}{
   \parbox[t]{\textwidth}{
       \setlength{\baselineskip}{0.5\baselineskip}
       $^a$The new equivalent width of this line in 51~Peg is 15.2\,m\AA\ smaller than EAGLNT's measurement.  This poor
agreement is caused by a nearby weaker unidentified line at 8772.54\,\AA, which at the 60\,000 resolution of the
new observations is partially resolved from the Al\,{\sc i} line and is excluded from the new equivalent width, but at
the lower 30\,000 resolution used by EAGLNT for 51~Peg was blended with the Al\,{\sc i} line and was included in their
equivalent width.  In the Sun this nearby line is excluded from both the new and EAGLNT equivalent widths
                         }
                   }
\end{tabular}
\end{flushleft}
\end{table*}

For all three Na\,{\sc i} lines used in the
new analysis the average [Na/H] is $+0.26 \pm 0.01$ (standard deviation
of the mean) which is very similar to the new analysis's value of 
$+0.25 \pm 0.01$ for the two lines in common with EAGLNT's analysis,
while for all seven Al\,{\sc i} lines used in the new analysis the
[Al/H] is $+0.21 \pm 0.01$ which is the same as the new analysis's value of
$+0.21 \pm 0.00$ for the two lines in common with EAGLNT's
analysis.  The additional Na\,{\sc i} and Al\,{\sc i} lines
of the new analysis thus serve to confirm the abundances derived
from the Na\,{\sc i} and Al\,{\sc i} lines in common with
EAGLNT's analysis.

EAGLNT's Na, Mg and Al abundances were based on only two lines
for each element and one of the Mg lines (Mg\,{\sc i} 8712.69\,\AA)
and one of the Al lines (Al\,{\sc i} 8772.87\,\AA) are not of top quality
\footnote{The 8712.69\,\AA\ Mg\,{\sc i} line is close to a stronger
Fe\,{\sc i} line at 8713.21\,\AA; it is not used here.
The 8772.87\,\AA\ Al\,{\sc i} line is partially
blended with a much weaker, but not insignificant, unidentified
line at 8772.54\,\AA.  This Al\,{\sc i} line, which at the resolution (60\,000)
of the present investigation can be reliably measured in
sharp-lined stars, is used here (see Table~2).}
so their abundances for these elements were less accurate than for
elements with good spectral representation such as Fe or Ni.
Overinterpretation of their results for Na, Mg and Al, thus,
tempted them to see overabundances of these three elements where
none actually existed.  The preponderance of NaMgAl stars
they observed at McDonald Observatory compared with ESO
is consistent with such an explanation.
The 189 stars in EAGLNT were made up
of 102 northern stars observed from McDonald Observatory at a
resolution of 30\,000, 71 southern stars observed from ESO at
resolutions of 60\,000 and 80\,000 and 16 stars observed from both
observatories.  Of the eight suggested NaMgAl stars
six were observed from McDonald Observatory and only two were
observed from ESO.  The disproportionately large number of
``detections'' among their McDonald stars can be attributed to
the larger observational scatter of the abundances, derived from
the lower resolution McDonald spectra.

Inspection of Fig.~5 and 6 shows nothing unusual about 51~Peg
compared to the other stars in these plots and so,
although 51~Peg is metal rich, we find no sign of its
composition being distinctive compared to other metal rich stars
of similar spectral type.

Finally we briefly discuss our new determination of 51~Peg's
metallicity.  Our result, [Fe/H] $= 0.20 \pm 0.07$, 
is higher than the earlier result of EAGLNT
who found [Fe/H] $= 0.06 \pm 0.07$ and is almost identical
to the results of Valenti (1994) and Gonzalez (1997) who found
[Fe/H] of $0.19 \pm 0.05$ and $0.21 \pm 0.05$, respectively.
Both Valenti's and Gonzalez's studies are also based on
high-resolution spectroscopic observations.  Most of the
increase of our [Fe/H] compared with EAGLNT's is a consequence
of our new equivalent widths, with smaller increases being caused by
the higher effective temperature ($\Delta T_{\rm eff} = +20$\,K)
and lower microturbulence ($\Delta \xi_{\rm t} = -0.22$)
used in the new analysis.
Valenti and Gonzalez both use very similar model atmosphere
parameters for 51~Peg to ours; Valenti adopts 
$T_{\rm eff} = 5724$\,K,
$\log g = 4.30$ and
$\xi_{\rm t} = 0.93$\,km\,s$^{-1}$
and Gonzalez adopts
$T_{\rm eff} = 5750$\,K,
$\log g = 4.40$ and
$\xi_{\rm t} = 1.0$\,km\,s$^{-1}$,
while we adopt (Table~1)
$T_{\rm eff} = 5775$\,K,
$\log g = 4.35$ and
$\xi_{\rm t} = 1.25$\,km\,s$^{-1}$.


   \section{Conclusions}

Our new study does not confirm the Na, Mg and Al overabundances
EAGLNT claimed to find.
With the aid of new spectra, that for most stars have higher spectral
resolution and include more Na, Mg and Al lines than EAGLNT were able to use,
we find that the
Na, Mg and Al abundances of the NaMgAl stars are the same
to within $\sim 0.03$\,dex as those
of non-NaMgAl stars of similar temperature, gravity and
metallicity.  The group thus appears to be illusory and so should
be deleted from the inventory of chemically peculiar stars.

The demise of the NaMgAl group of stars removes the
basis for EAGLNT's speculation that these stars,
and 51~Peg in particular, may have stellar remnant companions.
The identification of 51~Peg's companion as a planet, thus,
becomes more secure.  We note, however, Gray's (1997)
recent claim that the regular 4.23\,d radial velocity variation
of 51~Peg is of pulsational origin, rather than a result of
reflex orbital motion.  Clearly this question must be settled
before one can be sure that 51~Peg really has a planetary
companion.


\begin{acknowledgements}
We thank Al Wootten for pointing out that
51~Peg was one of the stars in EAGLNT's
survey, Brian Marsden for providing an ephemeris for Iris and
Paul Barklem and Jim O'Mara for calculating and
sending us line broadening data before publication.
DLL and JT acknowledge support from the US National Science
Foundation (grant AST 93-15124) and the Robert A. Welch Foundation.
BE and BG acknowledge support from the Swedish Natural Sciences Research
Council.
\end{acknowledgements}



\begin{thebibliography}{}

   \bibitem{} Anstee S.D., O'Mara B.J., 1995, MNRAS 276, 859

   \bibitem{} Barklem P.S., O'Mara B.J., 1997, MNRAS, submitted

   \bibitem{} Butler R.P., Marcy, G.W., Williams E., Hauser H., Shirts P.,
    1997, ApJ 474, L115

   \bibitem{} Edvardsson B., 1988, A\&A 190, 148

   \bibitem{} Edvardsson B., Andersen J., Gustafsson B., Lambert D.L.,
    Nissen P.E., Tomkin J., 1993, A\&A 275, 101 (EAGLNT)

   \bibitem{} Feltzing S., Gustafsson B., 1997, A\&A, submitted

   \bibitem{} Gonzalez G., 1997, PASP, submitted

   \bibitem{} Gray D.F., 1997, Nature 385, 795

   \bibitem{} Grevesse N., Noels A., Sauval A.J., 1996, in ASP Conf. Ser. 99,
     eds. S.S. Holt, G. Sonneborn, p.\ 117

   \bibitem{} Gustafsson B., Bell R.A., Eriksson K., Nordlund \AA, 1975,
    A\&A 42, 407

   \bibitem{} Kurucz R.L., Furenlid I., Brault J., Testerman L., 1984,
    Solar Flux Atlas from 296 to 1300\,nm, National Solar Observatory,
    Sunspot, New Mexico

   \bibitem{} Magain P., 1984, A\&A 134, 189

   \bibitem{} Marcy G.W., Butler R.P., Williams E., et al.,
    1997, ApJ, submitted

   \bibitem{} Mayor M., Queloz D., 1995, Nature 378, 355

   \bibitem{} Moore C.E., Minnaert M.G.J., Houtgast J., 1966, The Solar
    Spectrum 2935\,\AA\ to 8770\,\AA, National Bureau of Standards,
    Monograph 61

   \bibitem{} Piskunov N.P., Kupka F., Ryabchikova T.A., Weiss W.W.,
    Jeffery C.S., 1995, A\&AS 112, 525

   \bibitem{} Tull R.G., MacQueen P.J., Sneden C., Lambert D.L., 1995,
    PASP 107, 251

   \bibitem{} Valenti J.A., 1994, PhD thesis, Univ. of Colorado

\end{thebibliography}
\end{document}